\documentclass[%
 reprint,
superscriptaddress,
nofootinbib,
 amsmath,amssymb,
 aps,
prd,
]{revtex4-2}

\usepackage[dvipsnames]{xcolor}
\usepackage{booktabs}
\usepackage{tabularx}
\usepackage{amssymb,amsmath,amsfonts}
\usepackage{dsfont} 
\usepackage{graphicx,bm}
\usepackage[hidelinks,colorlinks=true,linkcolor=blue,citecolor=blue]{hyperref}

\newcommand\mnras{MNRAS}             

\usepackage{tabu}
\usepackage{array}
\usepackage{multirow}
\usepackage{CJK}
\usepackage{mathrsfs}
\usepackage{dcolumn}
\usepackage{mathtools}
\usepackage{graphicx}
\usepackage{dcolumn}
\usepackage{bm}
\usepackage[normalem]{ulem}

\begin{document}

\title{Orbital eccentricity can make neutron star g-mode resonances observable with current gravitational-wave detectors}

\author{János Takátsy}
\email{janos.takatsy@uni-potsdam.de}
\affiliation{Institut für Physik und Astronomie, Universität Potsdam, Haus 28, Karl-Liebknecht-Str. 24-25, Potsdam, Germany}
\affiliation{Niels Bohr International Academy, The Niels Bohr Institute, Blegdamsvej 17, DK-2100, Copenhagen, Denmark}

\author{Lorenz Zwick}
\affiliation{Niels Bohr International Academy, The Niels Bohr Institute, Blegdamsvej 17, DK-2100, Copenhagen, Denmark}
\affiliation{Center of Gravity, Niels Bohr Institute, Blegdamsvej 17, 2100 Copenhagen, Denmark}
\author{Pankaj Saini}
\affiliation{Niels Bohr International Academy, The Niels Bohr Institute, Blegdamsvej 17, DK-2100, Copenhagen, Denmark}
\affiliation{Center of Gravity, Niels Bohr Institute, Blegdamsvej 17, 2100 Copenhagen, Denmark}
\author{Johan Samsing}
\affiliation{Niels Bohr International Academy, The Niels Bohr Institute, Blegdamsvej 17, DK-2100, Copenhagen, Denmark}
\affiliation{Center of Gravity, Niels Bohr Institute, Blegdamsvej 17, 2100 Copenhagen, Denmark}


%



\newcommand{\cyan}{\color{cyan}}
\newcommand{\PS}[1]{{\cyan [PS: #1]}}             

\newcommand{\blue}{\color{blue}}
\newcommand{\JS}[1]{{\blue [JS: #1]}}

\newcommand{\dandelion}{\color{Dandelion}}
\newcommand{\JT}[1]{{\dandelion [JT: #1]}}               

\begin{abstract}
Dynamical tides can provide us vital information about the properties of neutron star (NS) matter. This is particularly true for g-modes, whose frequency and tidal coupling are highly sensitive to the composition of NSs, especially in their centers, where microphysical models are the least reliable. However, due to their weak coupling to external tidal fields, their effect on the gravitational-wave (GW) signal of binary inspirals can be difficult to observe. Here we show that the detectability of these tides can be significantly enhanced by binary NSs with moderate eccentricities. This is primarily due to higher eccentric harmonics in the early phase of the binary evolution experiencing larger phase shifts, which they transport to the sensitive band of GW detectors. In addition, g-mode tides in eccentric binaries undergo several epicyclic resonances, which also amplify the total phase shift. We demonstrate that these effects increase the detectability of g-mode dynamical tides by more than an order of magnitude for eccentricities of $e_\mathrm{10Hz}\sim0.2-0.4$, making it possible to put robust constraints on g-mode properties using current GW detectors, while all relevant models could potentially be constrained with eccentric binary NSs with Einstein Telescope.
\end{abstract}

\maketitle

\section{Introduction}
\label{sec:introduction}

The observation of gravitational waves (GWs) from a binary neutron star (NS) merger, GW170817, and subsequent electromagnetic observations opened a new chapter in multi-messenger astronomy \cite{LIGOScientific:2017vwq,LIGOScientific:2017ync,LIGOScientific:2017zic}, conveying valuable information about matter under extreme conditions inside NSs. While the LIGO-Virgo-Kagra (LVK) Collaboration has observed hundreds of GW transients from binary inspirals -- most of them produced by binary black holes (BHs) --, a second binary NS merger with electromagnetic counterpart is yet to be observed \cite{LIGOScientific:2025slb}. Nonetheless, this event alone, combined with other electromagnetic observations, like mass measurements of NSs in binary systems \cite{Demorest2010,Antoniadis2013,Cromartie2019,NANOGrav:2017wvv,Saffer:2024tlb,Romani:2021xmb,Romani:2022jhd,Romani:2025ytn} or X-ray pulse-profile modeling of various pulsars \cite{Riley:2019yda,Miller:2019cac,Riley:2021pdl,Miller:2021qha,Salmi2024,Salmi2024b,Vinciguerra2024,Dittmann2024,Choudhury2024,Qi2025,Mauviard2025,Miller2025}, already enabled putting stringent constraints on the equation of state (EoS) of dense nuclear matter \cite{De:2018uhw, Tews:2018iwm,Rezzolla2017,Annala:2017llu,Annala:2021gom,Pang:2021jta,Malik:2022zol,Altiparmak:2022bke,Marczenko:2022jhl,Takatsy:2023xzf,Koehn:2024set}.

Determining the state of nuclear matter at high densities, governed by the strong interaction, is a notoriously difficult problem. Even lattice Quantum Chromodynamics (QCD) methods fail to converge at such densities due to the notorious sign problem \cite[e.g.][]{Guenther2020,Attanasio2020,Borsanyi2021}. While at low densities, below nuclear saturation, the EoS is well established based on experimental data \cite[e.g.][]{Akmal1998,Wiringa1988}, hadron resonance gas models, as well as Chiral Effective Field Theory \cite{Huovinen2009,Bazavov2012,Lynn2015,Tews2018}, according to state-of-the-art calculations, the uncertainties of the nuclear EoS above $\sim1.1n_0$ (with $n_0$ the nuclear saturation density) become increasingly significant \cite{Tews2012,Tews2019}. Perturbative QCD also becomes an efficient tool at high densities ($\gtrsim40n_0$), where quarks become asymptotically free \cite{Gorda2018,Gorda2021,Gorda2023}. At densities between these regions, however, the EoS becomes highly uncertain. Even though one can rely on approximate global symmetries of QCD to estimate the EoS and the position of phase transitions \cite[e.g.][]{Kovacs:2021ger}, these transitions can virtually happen at any density. This makes the observation of NSs, the matter of which lies in this density regime an essential tool for studying the strong interaction \cite[e.g.][]{Annala:2017llu,Dietrich:2020efo,Kovacs:2021ger,Huth:2021bsp,Marczenko:2022jhl,Takatsy:2023xzf}.

The observation of GW170817 provided especially robust constraint on the EoS through the measurement of tidal deformability, which forms an integral part of GW modeling of binary NSs. In addition to adiabatic tides, state-of-the art models also include contributions from dynamical tides \cite{Schmidt:2019wrl,Haberland:2025luz,Abac:2023ujg}. These resonant effects, however, only contribute meaningfully to the waveform during the last couple of orbits, due to the resonant frequency of NS f-modes being several times larger than the orbital frequency of binary NSs near merger. This likely puts the first measurement of dynamical tides to the era of next-generation GW detectors, such as the Einstein Telescope (ET) \cite{ET:2025xjr}, even though their omission might bias parameter estimation already during LVK's fifth observing run \cite{Pratten:2021pro}. Additionally, non-negligible orbital eccentricities might help enhancing these effect \cite{Vick:2019cun,Yang:2018bzx,Yang:2019kmf,Takatsy:2024sin}.

In addition to f-modes, the properties of which mainly depend on bulk properties of NSs \cite{Chan:2014kua,Godzieba:2020bbz,Pradhan:2022vdf,Pradhan:2023zmg}, such as their mass and radius, NS seismology also predicts other types of oscillation modes, which connect more strongly to the precise nuclear physics above crust densities in the expense of having a weaker coupling to external driving forces. One type of these modes, the so-called gravity-modes (g-modes) have been studied extensively from their dependence on the EoS \cite[e.g.][]{Kantor:2014lja,Passamonti:2015oia,Constantinou:2021hba,Jaikumar:2021jbw,Zhao:2022toc,Counsell:2024pua} to their effect on GW signals and potential detectability \cite[e.g.][]{Lai:1993di,Ho:1998hq,Xu:2017hqo,Andersson:2017iav,Ho:2023shr,Pnigouras:2025muo}. The frequencies of the g-modes are closely tied to variations in the lepton fraction with density, or the presence of superfluidity \cite[e.g.][]{Kantor:2014lja,Passamonti:2015oia}, making them a powerful tool in telling different phases of nuclear matter apart \cite{Constantinou:2021hba,Jaikumar:2021jbw,Zhao:2022toc}. In addition, multiple recent studies suggest the emergence of interface-modes in NSs with a strong first-order phase transition, which create discontinuities in the density profile of NSs \cite{Miniutti:2002bh,Counsell:2025hcv,Zhu:2022pja,Lau:2020bfq,Rodriguez:2025oes}. However, due to their weak coupling to external tides, these modes might also potentially only become observable in the ET era for circular binaries.

\begin{figure}[!t]
    \centering
    \includegraphics[width=0.45\textwidth]{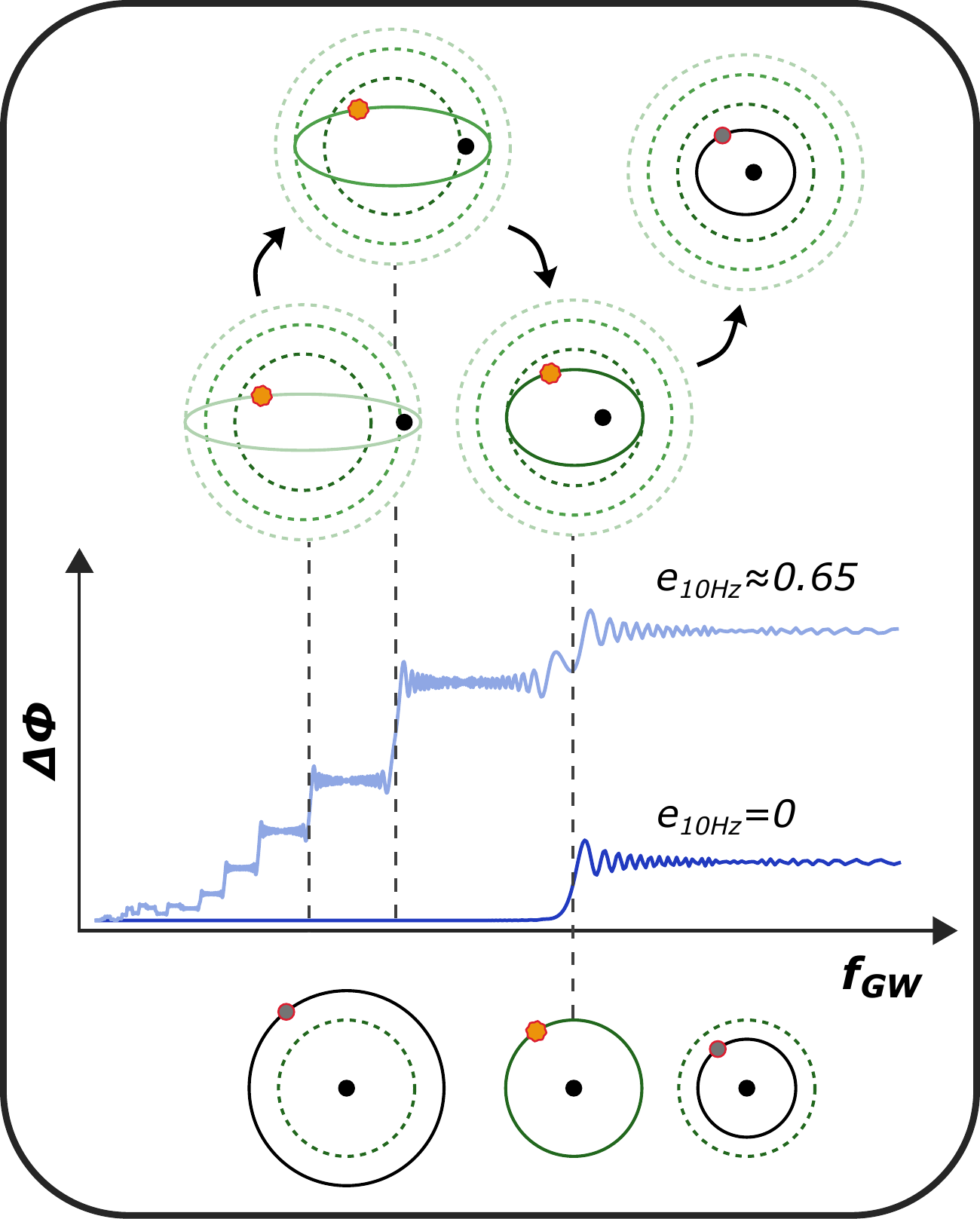}
    \caption{Illustration of the effect of g-mode resonances in circular and eccentric binaries, considering a single g-mode. Here, the orbital phase shift due to resonances is shown as a function of the $\ell=2$ mode GW frequency. While in a circular binary the quadrupolar tides only resonate with the $\ell=2$ harmonic of the binary, in the eccentric case the total phase shift of the binary will be accumulated through several epicyclic resonances (shown by the orange NSs). These resonances occur at well-defined orbital frequencies (denoted by dashed lines), whenever the frequency of the tidal mode becomes an integer multiple of the orbital frequency.}
    \label{fig:Illustration}
\end{figure}

In this paper we propose that orbital eccentricity could enhance the effect of g-mode resonances on the orbital evolution, potentially opening up the possibility for detections by current ground-based GW detectors. Even though circular binary inspirals are expected to make up the bulk of GW observations, measurable eccentricities are also expected to be produced for compact object binaries in dense stellar clusters during dynamical interactions like single-single gravitational-wave capture \cite{OLeary:2008myb,Kocsis:2011dr,Gondan:2017wzd,Takatsy:2018euo,Rasskazov:2019gjw,Gondan:2020svr,Zevin:2021rtf}, binary–single \cite{Samsing:2013kua,Samsing:2017xmd} or binary–binary interactions \cite{Zevin:2018kzq,Arca-Sedda+2021}, the secular evolution of hierarchical triple systems \cite{Wen:2002km,Antonini:2012ad,Antognini:2013lpa,Silsbee:2016djf,Hoang:2017fvh,Petrovich_Antonini2017,Rodriguez:2018jqu,Fragione:2019hqt,Fragione+2019}, or interactions in gaseous disks \cite{Tagawa:2020jnc,Samsing:2020tda,Fabj_Samsing2024,Li+2023,Rowan+2023,Whitehead:2023hmh,Rowan:2025xxb,Wang:2025xjy}. In fact, multiple recent GW observations show significant support for non-negligible eccentricities at frequencies inside the sensitive bands of LVK detectors \cite{Gamba:2021gap,Gayathri+2022,Romero-Shaw:2022fbf,Romero-Shaw:2022xko,Gupte:2024jfe,Iglesias:2022xfc,Romero-Shaw:2025vbc}, one of them, GW200105, being a NS-BH binary with a possible triple star origin \cite{Morras:2025xfu,Jan:2025fps,Stegmann:2025clo,Phukon:2025cky,Kacanja:2025kpr}. The eccentricity of the binary corresponding to GW200105 was inferred to be $0.1-0.15$ at a GW frequency of 20~Hz \cite{Morras:2025nlp,Jan:2025fps,Kacanja:2025kpr}. If binary NSs have residual eccentricities at orbital frequencies similar to g-mode frequencies, then these g-modes will be resonantly excited not just at single orbital frequency (as in the case of circular binaries), but by all the higher harmonics present in the spectrum of the eccentric system (see Fig.~\ref{fig:Illustration}). Additionally, if not accounted for, even small residual eccentricities, $\sim (10^{-3}\mbox{--}10^{-4})$ at $10$~Hz, can bias the NS EoS in current and third-generation detectors \cite{Favata:2013rwa, DuttaRoy:2024aew}. While the measurement of eccentricities in GW signals is challenging, there has been significant improvement in providing accurate waveform models \cite{Memmesheimer:2004cv,Huerta:2016rwp,Ramos-Buades:2021adz,Islam:2024zqo,Morras:2025nlp, Cao:2017ndf,Liu:2019jpg}, state-of-the-art effective-one-body models capable of accurately modeling eccentric waveform for systems with $e\lesssim 0.5$ and aligned spins \cite{Gamboa:2024imd,Gamboa:2024hli}. Due to these recent improvements and further significant efforts being currently made for pushing this eccentricity limit up to $e\sim0.9$ make this study especially timely.

This paper is structured as follows. In Sec.~\ref{sec:methods} we describe the Newtonian model used to describe tidal interactions, derive the formulas within this model for the energy exchange during epicyclic resonances, calculate the total phase shift inflicted on the GW signal of eccentric binaries due to consecutive resonances, and describe the eccentric waveform model we utilized for assessing the detectability of these phase shifts. In Sec.~\ref{sec:results}, after confirming the validity of our analytic formulas with direct numerical integration, we show how the detectability of g-mode resonances is significantly increased for binaries with finite orbital eccentricities, and show the extent this could improve on constraints based on circular binaries. Finally in Sec.~\ref{sec:conclusion} we offer our conclusions and discuss potential caveats.

\section{Methodology}
\label{sec:methods}

\subsection{Tidal model}

We use the notation of Ref.~\cite{Flanagan:2007ix} to model the tidal interaction of NS binaries. We focus only on quadrupole tides as they are dominant over higher-order multipoles. We thus label separate modes with $\alpha=\{n,l=2,m\}$. Additionally, we will later restrict our discussion to $n=1$ g-modes, which are expected to have the largest phase shift contribution among low-frequency modes for non-spinning NSs without strong phase transitions. The binary components have the masses $m_1$ and $m_2$, with $m_1$ being the primary mass, and tidal deformabilities corresponding to different modes $\lambda^\alpha_1$ and $\lambda^\alpha_2$. For simplicity, the equations presented here correspond to $\lambda^\alpha_2 = 0$, however, tidal interactions on the secondary are also taken into account in our results, and can easily be re-introduced by adding tidal terms with an exchange of indices $1\leftrightarrow2$. Then, the Lagrangian of the binary system with quadrupole tidal interactions reads:
\begin{eqnarray}
L &=&  \left[{1 \over 2} \mu {\dot r}^2 + {1 \over 2} \mu
r^2 {\dot
    \varphi}^2 + { m_\mathrm{tot} \mu \over r} \right] -  \sum_\alpha{1 \over 2} Q^\alpha_{ij} {\cal
    E}_{ij} \nonumber \\
&& + \sum_\alpha{1 \over 4 \lambda_{1}^\alpha \omega_\alpha^2} \left[ {\dot Q}^\alpha_{ij} {\dot
    Q}^\alpha_{ij} - \omega_\alpha^2 Q^\alpha_{ij} Q^\alpha_{ij} \right],
\label{eq:lagrange}
\end{eqnarray}
where $m_\mathrm{tot}=m_1+m_2$ and $\mu = m_1m_2/m_\mathrm{tot}$ are the total and reduced masses, and $\omega_\alpha$ is the frequency of the specific tidal mode. The Lagrangian is expressed in the Newtonian effective one-body form, with $r=|\boldsymbol{r}|\equiv|\boldsymbol{x}_1-\boldsymbol{x}_2|$ being the relative distance of the two objects and $\varphi$ the true longitude, i.e. the angle between the unit vector $\boldsymbol{e}_x$ and $\boldsymbol{r}$. The motion of the binary is also restricted to the $x-y$ plane.

Traditionally the tidal couplings are expressed in terms of tidal overlap integrals $\mathcal{Q}_\alpha$ \cite[se e.g.][]{Lai:1993di}, however, we prefer the use of tidal deformabilities, which can be directly connected to other results in the GW literature, especially those derived for f-mode tides. The connection between these two for non-spinning NSs is given by:
\begin{equation}
    k_\alpha = \frac{2\pi}{2l+1} \frac{\tilde{\mathcal{Q}}_\alpha^2}{\tilde{\omega}_\alpha^2} \: ,
\end{equation}
where $l=2$ for quadrupole tides, $k_\alpha = 3/2\cdot\lambda_\alpha R^{-5}$, with $R$ being the radius of the NS, is the contribution of a specific mode to the tidal Love number of the NS, and $\tilde{\mathcal{Q}}_\alpha = \mathcal{Q}_\alpha/(mR^l)$ and $\tilde{\omega}^2 = \omega^2 \left( Gm/R^3 \right)^{-1}$ are the dimensionless overlap integrals, and mode frequencies, respectively. For the highest frequency g-modes (with a single radial node) microphysical calculations of $\tilde{\mathcal{Q}}$ and $\tilde{\omega}$ give us dimensionless tidal deformabilities $\Lambda = \lambda_1/m^5$ in the range $10^{-2}-1$ \cite[e.g.][]{Counsell:2024pua,Lai:1993di,Yu:2016ltf}. Therefore, in this paper we choose $\Lambda_g=0.1$ as the typical value for the tidal deformability associated with the first g-mode.

The components of the external tidal field in Eq.~\eqref{eq:lagrange} can be expressed as
\begin{eqnarray}
     {\cal E}_{ij} &=&-m_2 \partial_i \partial_j \left( {1 \over r} \right) = -{m_2 \over r^3}(3n^in^j-\delta^{ij}) = \nonumber \\
     &=& -{m_2 \over r^3}\begin{pmatrix}
    {1\over2}+{3\over2}\cos 2\varphi&{3\over2}\sin 2\varphi&0\\
    {3\over2}\sin 2\varphi&{1\over2}-{3\over2}\cos 2\varphi&0\\
    0&0&-1
    \end{pmatrix}.
\end{eqnarray}
Due to the symmetrical and traceless nature of the tensors $\boldsymbol{Q}_\alpha$ and $\boldsymbol{\mathcal{E}}$, their five independent components can be represented by spherical-harmonic tensors, $\boldsymbol{\mathcal{Y}}_{lm}$, as in Ref.~\cite{Takatsy:2024sin}. Their expansion in this basis then reads:
\begin{equation}
\boldsymbol{Q}_\alpha=\sum_m Q^\alpha_m \boldsymbol{\mathcal{Y}}_{2m}\,,
\quad
\boldsymbol{\mathcal{E}} = \sum_m \mathcal{E}_m \boldsymbol{\mathcal{Y}}_{2m}\,.\quad
\end{equation}
The coefficients for the external tide then become \cite[see][]{Takatsy:2024sin}:
\begin{equation}
    \mathcal{E}_0 = -\sqrt{3 \over 2}{m_2 \over r^3}\,,\quad
    \mathcal{E}_{\pm2} = -{3 \over 2}{m_2 \over r^3} e^{\mp i2\varphi}\,,\quad
    \mathcal{E}_{\pm1} = 0 \ ,
\end{equation}
where the $m=\pm 1$ components are zero because the orbital motion is restricted to the $x-y$ plane.

From the Lagrangian in Eq.~\eqref{eq:lagrange} the equations of motion for the independent quadrupole components are
\begin{eqnarray}
    {\ddot Q}^\alpha_m + 2\gamma_\alpha {\dot Q}^\alpha_m + \omega_\alpha^2 Q^\alpha_m &=& -\lambda^\alpha_1 \omega_\alpha^2 \mathcal{E}_m \ ,
    \label{eq:EoM}
\end{eqnarray}
where we introduced a dissipative term proportional to $\gamma_\alpha$. This dissipative term can be either arise from microscopic viscosity, or back-reaction forces as well due to the GW radiation of the tidal bulge. Dissipative forces due to microscopic viscosity in old, and hence cold NSs is expected to be negligible \cite{Cutler1990,Kochanek:1992wk,Bildsten:1992my}. Moreover, while GW radiation exerts a non-negligible force on the system for high-frequency modes typical for f- and p-modes, its relevance drops sharply for lower frequencies, as the coefficient $\gamma_\alpha$ in this case scales with $\omega_\alpha^6$ of the respective mode. Hence going from $1.5-2$~kHz for f-modes to the characteristic frequency of $\mathcal{O}(100)$~Hz typical for g-modes, the dissipation coefficient drops by 6 orders of magnitude, and hence also become negligible. Therefore, in this study we assume $\gamma_\alpha\approx0$ and only consider orbital GW radiation as a source of dissipation.

Let us also mention that although in this study we assume a linear tidal model, some recent studies have pointed out that non-linearities might become relevant for resonant excitations of low-frequency tides and produce interesting phenomena, such as resonance-locking, which might further enhance the amplitude of the excited modes \cite{Kwon:2025zbc}. Since the aim of this study is to point out the importance of eccentricity in the observability of tides, we neglect non-linearities and leave the pursuit of these effects to follow-up studies.

\subsection{Epicyclic resonances}

While f-modes cannot be coherently excited throughout multiple orbits in eccentric binaries \cite[see e.g.][]{Vick:2019cun,Takatsy:2024sin}, for the typical frequency of $\mathcal{O}(100)$~Hz of g-modes, resonant excitations, lasting several orbital periods can occur, even with non-zero eccentricities. Analytic formulas determining the excited tidal amplitude during such resonances have traditionally been derived for circular binaries \cite[see e.g.][]{Lai:1993di,Pnigouras:2025muo}, and have been also shown for eccentric binaries as well in the context of f-modes \cite{Vick:2019cun}. However, due to the different notational convention, we reproduce these results here.

Let us then expand Eq.~\eqref{eq:EoM} in Fourier series. The right-hand side of Eq.~\eqref{eq:EoM}, i.e. the exciting force can be expressed as a sum over different harmonics (or epicycles):
\begin{equation}
    -\lambda^\alpha_1 \omega_\alpha^2 \mathcal{E}_m(t) = C_m \lambda^\alpha_1 \omega_\alpha^2 \frac{m_2}{a^3} \sum\limits_{\ell=-\infty}^{\infty} X_{-\ell}^{-3,-m} e^{-i \ell\Phi}\: ,
    \label{eq:E_X}
\end{equation}
where $a$ is the semi-major axis of the binary, $\Phi\equiv \Phi(t)$ is the mean anomaly, and $C_m = 3/2$ for $m=\pm2$ and $C_m=\sqrt{3/2}$ for $m=0$. Note the difference between $l$, which is related to the multipole expansion, and $\ell$, which is the index of different harmonics. $X_\ell^{k,m}$ are the so-called Hansen-coefficients, defined by
\begin{equation}
    X_\ell^{k,m}(e) = \frac{1}{2\pi} \int\limits_{-\pi}^{\pi} \left[ \frac{r(\Phi)}{a} \right]^k \cos(m\varphi - \ell\Phi) \mathrm{d}\Phi \ .
\end{equation}

As we are interested in the evolution of the amplitude of the tidal oscillations during resonances, let us express the temporal evolution of the tides the following way:
\begin{equation}
    Q_m^\alpha(t) = q_m^\alpha(t) e^{-i\cdot\mathrm{sgn}(m)\omega_\alpha t} \: ,
\end{equation}
where $q_m^\alpha(t)$ is the amplitude of the mode. Then neglecting $\ddot{q}_m^\alpha(t)$, we get
\begin{align}
    -2i\omega_\alpha \dot{q}_m^\alpha(t)e&^{-i\cdot \mathrm{sgn(m)}\omega_\alpha t} = \nonumber \\
    &C_m \lambda^\alpha_1 \omega_\alpha^2 \frac{m_2}{a^3} \sum\limits_{\ell=-\infty}^{\infty} X_{-\ell}^{-3,-m} e^{-i \ell\Phi} \: .
\end{align}
During a resonance one of the epicylic frequencies coincides with the natural frequency of the mode $\ell_\mathrm{r} \Omega = \mathrm{sgn}(m)\omega_\alpha$, where $\Omega$ is the orbital frequency of the binary. As the binary sweeps across the resonance, it excites the amplitude of the respective tide, giving:
\begin{equation}
    q_m^\alpha(t) = \frac{i \lambda_1^\alpha \omega_\alpha C_m}{2} \frac{m_2}{a^3}X_{-\ell_\mathrm{r}}^{-3,-m} \int\limits_{-\infty}^\infty \mathrm{d} t' e^{it'(\omega_\alpha - |\ell_\mathrm{r}|\Omega)} \: .
    \label{eq:q_res}
\end{equation}

Since the amplitude, $\mathcal{A}$, and frequency of the exciting force are slowly evolving, i.e. $\mathrm{d}(\ln \mathcal{A})/\mathrm{d}t \ll |\ell_\mathrm{r}|\Omega$, and $ |\ell_\mathrm{r}|\mathrm{d}\Omega/\mathrm{d}t \ll |\ell_\mathrm{r}|^2\Omega^2$, we can apply the stationary phase approximation, i.e. that at the resonance
\begin{equation}
    \frac{\mathrm{d}}{\mathrm{d}t}\left(\omega_\alpha t - |\ell_\mathrm{r}| \Omega t \right) = 0 \: .
\end{equation}
Then, since the first non-vanishing term in the series expansion of the exponent in Eq.~\eqref{eq:q_res} is quadratic, to first-order the integral reduces to a Gaussian integral, yielding:
\begin{equation}
    \left|q_m^\alpha\right| \approx \frac{C_m \lambda^\alpha_1 \omega_\alpha^2}{2} \frac{m_2}{a^3} X_{|\ell_\mathrm{r}|}^{-3,|m|} \sqrt{\frac{2\pi}{|\ell_\mathrm{r}|\dot{\Omega}_\mathrm{r}}} \: ,
\end{equation}
for $|m|=2$ and $q_0^\alpha=0$, where $\dot{\Omega}_\mathrm{r}\equiv \dot{\Omega}(\Omega=\omega_\alpha/|\ell_\mathrm{r}|)$ and where we have used the identity $X_{-\ell}^{k,-m}=X_{\ell}^{k,m}$.

Then, the change in the energy of the modes during the resonance is
\begin{align}
    \Delta E &= \Delta\left\{\frac{2}{4\lambda_1^\alpha \omega_\alpha^2} \left[ \left(\dot{Q}_2^\alpha\right)^2 + \omega_\alpha^2 \left(Q_2^\alpha\right)^2 \right]\right\} = \frac{1}{\lambda_1^\alpha} \left|q_2^\alpha\right|^2 \nonumber \\
    &= \frac{9\pi\lambda_1^\alpha\omega_\alpha^2}{8\ell_\mathrm{r}\dot{\Omega}_\mathrm{r}} \frac{m_2^2}{a^6} \left(X_{\ell_\mathrm{r}}^{-3,2}\right)^2 \: ,
\end{align}
where we have used that the $m=\pm 2$ modes obtain the same amplitude and $\ell_\mathrm{r}>0$, corresponding to the $m=2$ mode. Note that this formula only applies to NS modes that can be coherently excited throughout multiple orbits (see discussion in the following section).

\subsection{Dephasing in eccentric binaries}
\label{ssec:analytic_deph}

The energy- and angular momentum transfer between the tides and the orbit perturbs the orbital evolution, resulting in the gradual accumulation of dephasing in the GW signal. This will be due to both the direct change in the chirp of the signal due to the energy transfer, as well as the indirect effect of changing the relation between the frequency and eccentricity of the orbit, and thus modifying the leading-order GW radiation at a given frequency. These effects have been summarized in the recent work, Ref.~\cite{Takatsy:2025bfk}. Here we only concentrate on the direct dephasing due to energy transfer, since it is expected to give the dominant contribution, as the indirect effect is suppressed by positive powers of the eccentricity \cite{Takatsy:2025bfk}. Moreover, the aim of this paper is to show the potential enhancement in the detectability of g-mode resonances in eccentric binaries without aiming to derive an accurate prescription for the dephasing. However, we note that the indirect effect cannot be neglected for precise modeling of the waveform, especially for large eccentricities.

As it was shown by several studies \cite[see e.g.][]{Lai:1993di}, the phase shift due to changing the energy of the orbit by $-\Delta E$ due to small perturbations is
\begin{equation}
    \Delta \Phi \approx - \frac{3}{2}\frac{\Omega^2}{\dot{\Omega}} \frac{\Delta E}{\left| E_\mathrm{orb} \right|} \: ,
    \label{eq:dPhidE}
\end{equation}
where $E_\mathrm{orb}$ is the orbital energy. Note that this phase shift is negative for perturbations that extract energy from the orbit. Using the leading-order results of Ref.~\cite{Peters:1964zz}:
\begin{equation}
    \frac{\Omega}{\dot{\Omega}} = \frac{5}{96} \frac{a^4}{m_1^3(q+1)q \mathcal{F}(e)} \: ,
\end{equation}
where
\begin{align}
    \mathcal{F}(e) = \left(1+\frac{73}{24}e^2 + \frac{37}{96}e^4 \right)(1-e^2)^{-7/2} \: ,
    \label{eq:Fe}
\end{align}
and $q=m_2/m_1$ is the mass ratio. Then we get:
\begin{equation}
    \Delta \Phi_1 \approx -\frac{75\pi}{8192} \frac{\ell_\mathrm{r}\Lambda_1^\alpha}{q(q+1)} \left[ \frac{X_{\ell_\mathrm{r}}^{-3,2}}{\mathcal{F}(e_\mathrm{r})} \right]^2 \: ,
\end{equation}
where $\Lambda_1 = \lambda_1/m_1^5$. Then contributions from the g-mode resonances of the other object should also be added to get the total dephasing. For $m_1=m_2$ and $\Lambda^\alpha_1=\Lambda^\alpha_2 \equiv \Lambda_\alpha$ we get:
\begin{equation}
    \Delta\Phi \approx -\frac{75\pi}{8192} \ell_\mathrm{r}\Lambda_\alpha \left[ \frac{X_{\ell_\mathrm{r}}^{-3,2}}{\mathcal{F}(e_\mathrm{r})} \right]^2 \: .
\end{equation}

Resonances between the tidal modes and the various epicyclic frequencies can only take place if the orbital decay is sufficiently slow so that the excitation remains coherent for multiple orbits. This can be verified comparing the orbital period to the resonance time:
\begin{equation}
    \delta t_\mathrm{r} = \sqrt{\frac{2\pi}{\ell_\mathrm{r}\dot{\Omega}_\mathrm{r}}} \: ,
\end{equation}
which makes the condition for effective resonant excitations expressed through the number of orbits during a resonance the following:
\begin{align}
    \delta N_\mathrm{r} &= \frac{\Omega_\mathrm{r}}{2\pi}\delta t_\mathrm{r}  \gtrsim 1 \: ,
\end{align}
with
\begin{align}
    \delta N_\mathrm{r} = \left[ \frac{5}{192\pi} \frac{(1+q)^{1/3}}{q} \frac{\ell_\mathrm{r}^{2/3}}{(m_1\omega_\alpha)^{5/3} \mathcal{F}(e_\mathrm{r})} \right]^{1/2} \: .
\end{align}
It is also apparent from this equation that while g-mode tides with frequencies of $\mathcal{O}(100~\mathrm{Hz})$ can be resonantly excited, the f-mode tides of binaries the same eccentricity and $\ell_\mathrm{r}$ (meaning a higher orbital frequency) are much more difficult to bring into resonance due to their high frequency of $\mathcal{O}(1~\mathrm{kHz})$, suppressing $\delta N_\mathrm{r}$ by an order of magnitude.
\begin{figure}[!t]
    \centering
    \includegraphics[width=0.48\textwidth]{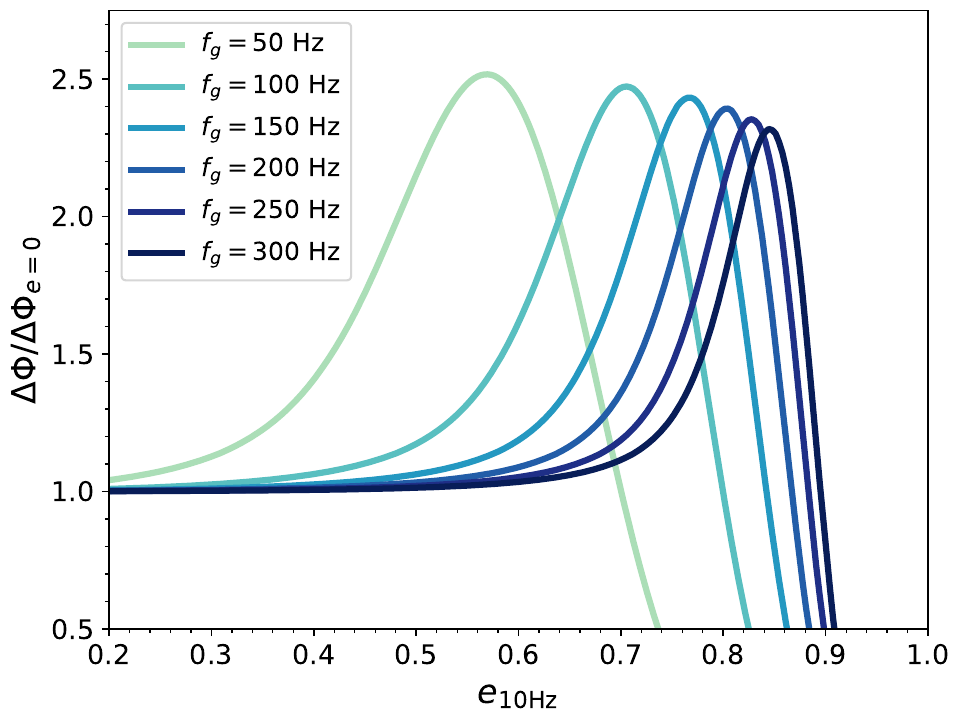}
    \caption{Time-domain phase shift, $\Delta\Phi$, normalized with $\Delta\Phi(e=0)$, as a function of orbital eccentricity at a reference GW frequency of $10$~Hz ($e_\mathrm{10Hz}$). Lines with different colors show results for different values of the g-mode frequency, $f_g$. The results in our parameterization are independent of the tidal coupling $\Lambda_g$.}
    \label{fig:DPhi}
\end{figure}

Fig.~\ref{fig:DPhi} shows the dependence of the total dephasing due to g-mode tides, normalized to the zero-eccentricity case, as a function of orbital eccentricity, $e_\mathrm{10Hz}\equiv e(f_\mathrm{GW}^{\ell=2}=10\,\mathrm{Hz})$. Note that the phase shift for circular binaries is independent of $f_g$ in the chosen parameterization. The maximal phase shift is $\sim2.5$ times larger than in the circular case, independent of the g-mode frequency, while the orbital eccentricity at which it is reached depends on $f_g$. For large eccentricities, the coherence condition, $\delta N_\mathrm{r}\gtrsim1$, is no longer satisfied, and hence the total dephasing due to g-mode resonances drops to zero.

To get the total phase shift then from tidal resonances in one of the objects, one has to sum up the contribution of the individual resonances for which the above condition is fulfilled:
\begin{equation}
    \Delta \Phi_1^\mathrm{tot} = \sum\limits_{\delta N_\mathrm{r}>1} \Delta\Phi_1(\ell_\mathrm{r},e_\mathrm{r}) \: .
\end{equation}
The formulas describing the phase shift from resonances in the second object are analogous.

In GW modeling often frequency-domain waveforms are being used. Then the time-domain phase shift derived above should be replaced by the corresponding frequency-domain phase shift, which, using the stationary-phase approximation becomes \cite{Cutler:1994ys,Takatsy:2025bfk}:
\begin{equation}
\label{eq:deph_simple}
    \Delta\Psi(F) = 2\pi F \Delta t(F) - \Delta\Phi(F) \: ,
\end{equation}
where $F$ is the orbital frequency and $\Delta t(F)$ is the time delay of the perturbed evolution reaching the orbital frequency $F$. The time delay generated by each resonance is simply given by:
\begin{equation}
    \Delta t_1 = \frac{\Delta \Phi_1}{\Omega_\mathrm{r}} \: .
\end{equation}
Thus the total frequency-domain phase shift becomes:
\begin{equation}
    \Delta \Psi_1^\mathrm{tot} = \sum\limits_{\delta N_\mathrm{r}>1} \left( \frac{F}{F_\mathrm{r}} -1 \right)\Delta\Phi_1 \: ,
\end{equation}
where $F_\mathrm{r}$ is the orbital frequency at the resonance point. The orbital phase shift is conventionally taken to be zero at the merger, and thus the phase shift is accumulated backwards in time. Then the total phase shift at a given frequency will be:
\begin{equation}
    \Delta\Psi_1^\mathrm{tot}(F) = \sum\limits_{\delta N_\mathrm{r}>1} \Delta\Phi_1 \left( 1-\frac{F}{F_\mathrm{r}} \right) \Theta(F_\mathrm{r} - F) \: ,
\end{equation}
where $\Theta$ is the Heaviside-function.

\subsection{Waveform model}
In this work, we employ a Newtonian waveform model that describes GWs for binaries at arbitrary eccentricity \cite{Mikoczi:2012qy}. In the stationary phase approximation \citep{Cutler:1994ys,Yunes:2009yz}, an eccentric waveform in the Fourier domain is given by the following sum of harmonics:
\begin{equation}
    \tilde{h}(f) \approx \sum_{\ell=1}^{\ell_\mathrm{max}} \tilde{\mathcal{A}}_\ell\left( f \right) \exp\left[ i\psi_\ell(f) \right],
\end{equation}
where $f=\ell F$ is the detector-frame GW frequency, $\psi_{\rm \ell}$ the phase of each harmonic, and we truncate the sum at $\ell_{\rm max}$. The coefficients $\tilde{\mathcal{A}}_{\ell}$ are a combination of Bessel functions and projection factors that also account for the observer position with respect to the binary's inclination and argument of pericenter. Here, we assume that the binary is optimally oriented and non-precessing. While precession, for example, cannot be generally neglected, especially for large eccentricities, we make this assumption in order to separate the effect of g-mode resonances from other types of orbital perturbations. We also note that while Newtonian waveforms are not sufficient for parameter inference, they are a useful tool to quantify the dependence of the signal-to-noise ratio (SNR) with eccentricity. Additionally, they have the advantage of being fast to evaluate, and having a definition of eccentricity that relates unequivocally to the geometric shape of of the orbit.

The value of $\ell_{\rm max}$ is mainly determined by modeling and observational considerations. Current eccentric effective-one-body waveforms are accurate up to $e\lesssim0.5$ corresponding to $\ell_\mathrm{max}\sim10-20$, while a limit of $e_\mathrm{max}\sim0.9$ is expected to be reached in the foreseeable future for spin-aligned configurations \cite{Gamboa:2024imd,Gamboa:2024hli}. Additionally, recovering significant power at large $\ell$ requires tracking repeated pericenter passages over a long observational baseline, simply because those modes correspond to the peak emission of a wide binary with a long inspiral timescale. In this work we adopt two benchmark choices. First, $e_{\rm max}=0.9$, consistent with the range expected to be supported by eccentric waveform models in the near future. Second, we explore the idealistic case with $\ell_{\rm max}\to\infty$, assuming exact modeling of the eccentric waveform up to arbitrary eccentricities. In the first case, $\ell_\mathrm{max}(f)$ is determined for a given GW frequency such that $e(F_{\ell_\mathrm{max}})\equiv e(f/\ell_\mathrm{max})\leq0.9$. This is roughly equivalent to setting $\ell_\mathrm{max}=80$ for all frequencies. We also note that this choice corresponds to a maximal required observation duration of at most $\sim$ a week (for $e_\mathrm{10Hz}\geq 0.1$), comfortably within detector capabilities.

The phases $\psi_{\ell}$ are given in \cite{Mikoczi:2012qy}, and describe the phase evolution of each eccentric harmonic. As detailed in \cite{Takatsy:2025bfk} and \citep{Zwick:2025qzv} they can be supplied with dephasing prescriptions related to orbital perturbations. The dephasing prescription detailed in Eq.~\eqref{eq:deph_simple} enters the various harmonics of the waveform in the following way:
\begin{align}
    \label{eq:dephasingboos}
    &\Delta \psi_{\ell}(f) = \ell\Delta \Psi\left(\frac{f}{\ell}\right),
\end{align}
where we used the fact that $f = \ell F$. This formula comes from the fact that each harmonic contributes power at the detector frequency $f$ only when the
binary is orbiting at $F = f/\ell$. In other words, different harmonics are effectively sampling different epochs of the inspiral: Higher $\ell$ modes trace the system at earlier times, when the
binary is at larger separations and evolving at lower orbital frequencies. Thus, although the sum of all harmonics are observed at the same Fourier frequency, the phase of each harmonic corresponds to different orbital frequencies. This mapping is the core reason higher harmonics experience stronger dephasing \citep{Takatsy:2025bfk,Zwick:2025qzv}.

Given a waveform and a way to prescribe dephasing, we use the $\delta$SNR criterion \citep[e.g.][]{Kocsis:2011dr,Zwick:2022dih} to assess whether the resonances produce a detectable effect. For a circular waveform:
\begin{align}
    \delta{\rm SNR}^2 = 4 \int \frac{|\delta \tilde h(f)|^2}{S_n(f)}\,df ,
\end{align}
where $S_n(f)$ is the noise power spectral density of the detector and we require $\delta{\rm SNR} > \mathcal{C}$. Here, the waveform difference is:
\begin{align}
    \delta\tilde h(f)
    &= \tilde h_{\rm tot}(f) - \tilde h_{\rm vac}(f),
\end{align}
where $\tilde h_{\rm tot}$ includes the dephasing and $\tilde h_{\rm vac}$ is the vacuum template. For eccentric systems, the criterion generalizes to a sum:
\begin{align}
    \delta{\rm SNR}^2
    &= 4 \int \frac{1}{S_n(f)}
       \left|\sum_{\ell} \delta\tilde h_\ell(f)\right|^2 \, df \nonumber \\
    &\approx 4 \int \frac{1}{S_n(f)}
       \sum_{\ell} \left|\delta\tilde h_\ell(f)\right|^2 \, df,
\label{eq:dSNR}
\end{align}
while keeping the same interpretation. Here we used the fact that the phases of different harmonics are incoherent and thus the frequency integral over mixed terms, $\delta\tilde{h}_\ell^* \delta\tilde{h}_{\ell'}$ with $\ell\neq\ell'$ becomes negligible. We take $\mathcal{C}=3$ and $\mathcal{C}=8$ as representative thresholds for marginal and confident detection. However, we note that this metric does not incorporate parameter degeneracies and therefore only serves as an indication whether these tidal effects could be measurable in the full parameter inference \citep[see e.g.][for Bayesian inference tests of environmental effects]{speri_acc,cole_nature,zwick_novel,garg_ecc, copparoni_stoch}. This is particularly important for highly eccentric systems, the dynamics of which are coupled to e.g. post-Newtonian spin effects \cite{will2014,isobel2022}.

\section{Results}
\label{sec:results}

\subsection{Numerical simulation}

In order to validate our analytic formulas we also performed direct numerical integrations of the equations of motion given by Eq.~\eqref{eq:EoM} for the tides, coupled to the evolution of the orbital elements, for which we included the 2.5PN gravitational-radiation reaction in addition to Newtonian forces. For the full set of equations see Appendix~B of Ref.~\cite{Takatsy:2024sin}. We then extracted the phase shift produced by dynamical tides using a numerically more convenient form of Eq.~\eqref{eq:dPhidE}, which gives the phase shift accumulated during a single orbit by:
\begin{equation}
    \Delta \Phi \approx - 2\pi \frac{\Delta E_Q}{\Delta E_\mathrm{GW}} \: ,
\end{equation}
where $\Delta E_Q$ and $\Delta E_\mathrm{GW}$ are the changes in the orbital energy during a single orbit due to tidal interactions and GW radiation, respectively.

\begin{figure}[!t]
    \centering
    \includegraphics[width=0.48\textwidth]{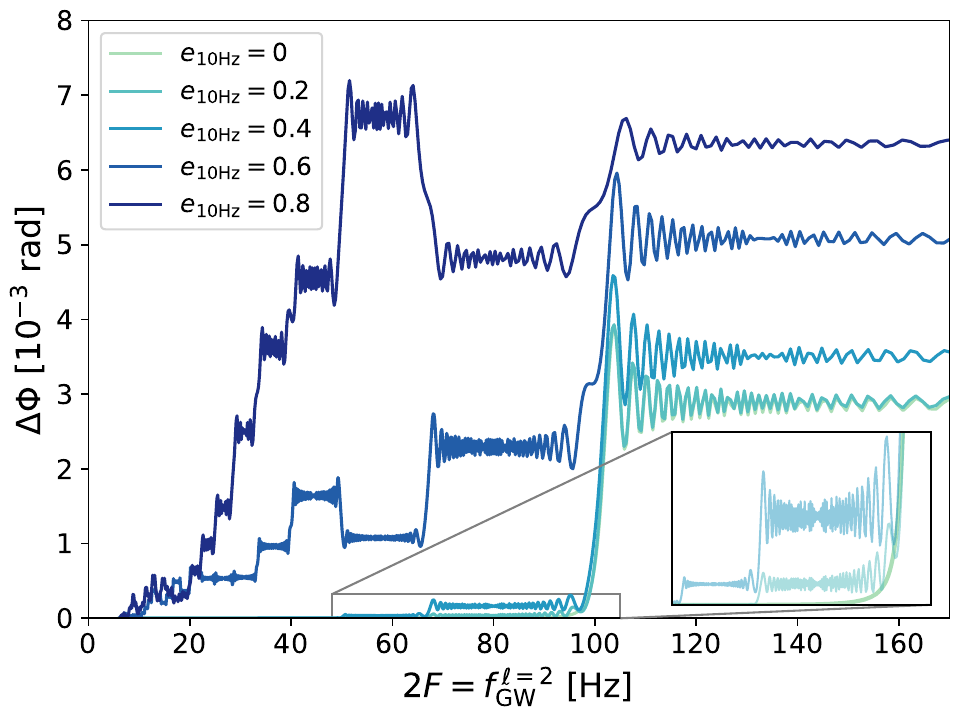}
    \caption{Time-domain phase shift, $\Delta\Phi$, of eccentric binary NSs due to g-mode resonances as a function of $2F = f_\mathrm{GW}^{\ell=2}$. Different lines show results for binaries with different orbital eccentricities at a reference GW frequency of $10$~Hz ($e_\mathrm{10Hz}$). The tidal parameters are chosen as $\Lambda_1^g=0.1$ and $\Lambda_2^g=0$ and the g-mode frequency is set to $f_g=100$~Hz.}
    \label{fig:Sim}
\end{figure}

The time-domain phase shift as a function of 2 times the orbital frequency is shown in Fig.~\ref{fig:Sim} for binary NSs of $m_1=m_2=1.4~M_\odot$ for several different orbital eccentricities at a reference GW frequency of $10$~Hz, i.e. $e_\mathrm{10Hz}\equiv e(f_\mathrm{GW}^{\ell=2}=10\,\mathrm{Hz})$. If one instead chose the GW peak frequency as a reference point, defined by \cite{Wen:2002km}:
\begin{equation}
    f_\mathrm{GW}^\mathrm{peak} \approx \frac{(1+e)^{1.1954}}{(1-e^2)^{3/2}}f_\mathrm{GW}^{\ell=2}  \: ,
\end{equation}
then the lowest eccentricities shown in Fig.~\ref{fig:Sim}, i.e. $e(f_\mathrm{GW}^{\ell=2}=10\,\mathrm{Hz}) = \{0, 0.2 \}$ would correspond to eccentricities $e(f_\mathrm{GW}^{\mathrm{peak}}=10\,\mathrm{Hz}) = \{0, 0.29 \}$, while all the higher eccentricities would translate to peak GW frequencies always higher than 10~Hz throughout the entire binary evolution.

The tidal parameters are chosen as $\Lambda_1^g=0.1$ and $\Lambda_2^g=0$ and the g-mode frequency is set to $f_g=100$~Hz. These will be our standard choice for tidal parameters, unless noted otherwise. In this representation, the circular mode resonance occurs at $f_\mathrm{GW}^{\ell=2}=f_g$, while it is also apparent that the resonances indeed occur at $f_\mathrm{GW}^{\ell=2}=2f_g/\ell$ for the general eccentric case. Note however, that these resonances do not necessarily add up constructively. This is because the phases of excitation during consecutive resonances are not coherent and thus a resonance can not just excite but de-excite previously established tidal oscillations. This creates a phase evolution that resembles a random walk along resonances. Note, however, that similarly to f-mode resonances in eccentric binaries \cite{Takatsy:2024sin}, the expectation value of the phase shift follows the simple sum outlined in Section~\ref{ssec:analytic_deph}, even though the total phase shift can be smaller or significantly larger than that. Also note that while, as expected, for $e=0$ only a single resonance occurs at $f_\mathrm{GW}=f_g$, only small contributions appear at lower frequencies for the $e=0.2$ and $e=0.4$ cases as well, while the total phase shift for $e=0.2$ almost completely overlaps with the zero-eccentricity case. As a final remark we note that the energy exchanges during resonances from our direct numerical integration agree with those given by the analytic formulas within a few percent.

\subsection{Detectability of dephasing}

\begin{figure}[!t]
    \centering
    \includegraphics[width=0.48\textwidth]{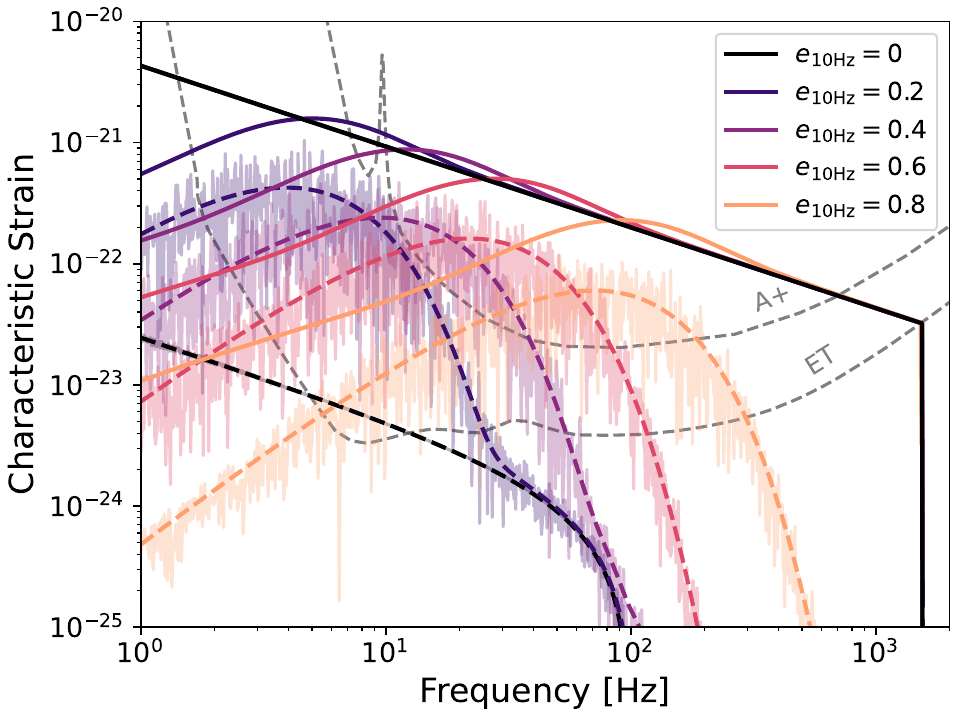}
    \caption{Characteristic strains, $h_c(f) = \tilde{h}(f)f^{1/2}$, of binary NSs at a distance of $40$~Mpc with different eccentricities at $f_\mathrm{GW}^{\ell=2}=10$~Hz (solid lines). Dashed lines correspond to the envelopes of the difference between the vacuum strains and strains from evolutions where g-mode resonances are taken into account, as defined by the second line in Eq.~\eqref{eq:dSNR}. The expectation values were used for calculating the phase shift. The parameters of the g-modes here are $f_\mathrm{g}=100$~Hz, and $\Lambda_1^g=0.1$ and $\Lambda_2^g=0$. The sensitivity curves of A+ and ET are also represented by gray dashed lines.}
    \label{fig:hc}
\end{figure}

Before examining the results for the $\delta$SNR, it is instructive to look at the the strains of the difference signals $\delta\tilde{h}(f)$. Several examples are shown in Fig.~\ref{fig:hc} for binaries with different eccentricities for our standard choice of tidal parameters at a distance of $40$~Mpc, corresponding to the distance of GW170817. The solid lines show the total strain of the vacuum waveforms, while the dashed curves correspond to the envelopes of the difference strains calculated from the sum of $|\delta\tilde h_\ell(f)|^2$ over all the harmonics. The difference strains containing all the phase information are also shown with semi-transparent lines. The waveforms are truncated at the last stable orbit, calculated as the solution of:
\begin{equation}
    \rho_\mathrm{p}(e_\mathrm{LSO}) = \frac{6 + 2e_\mathrm{LSO}}{1 + e_\mathrm{LSO}} \:,
\end{equation}
with $\rho_\mathrm{p}=a(1-e)/m_\mathrm{tot}$ the dimensionless pericenter distance and where we used Peters's formula to determine $a(e)$ \cite{Peters:1964zz}.

For the zero-eccentricity case we see that the the difference strain is zero above the the resonance frequency $f_g$, and $|\delta\tilde h(f)|/|\tilde h_\mathrm{vac}(f)|$ approaches a constant value for low frequencies, since the frequency-domain dephasing also tends to a constant. The situation becomes more interesting for the $e=0.2$ case. We saw in Figs.~\ref{fig:DPhi} and \ref{fig:Sim} that the change in the dephasing compared to the zero-eccentricity case is almost negligible for $e=0.2$, however, here we see a huge boost in the difference strain at GW frequencies of $\sim10$~Hz. This, as already discussed in Refs.~\cite{Takatsy:2025bfk,Zwick:2025qzv}, is due to the fact that as the binary becomes more eccentric at lower orbital frequencies, the orbital phase shifts are transported to higher frequencies by the various harmonics. More quantitatively, as the frequency-domain phase shift approaches $\Delta\Psi(F\approx0)\approx\Delta\Phi^\mathrm{tot}=\mathrm{const.}$, the phase shift of the $\ell^\mathrm{th}$ harmonic becomes $\Delta\psi_\ell(\ell F)\approx\ell\Delta\Phi^\mathrm{tot}$, which contributes to the difference strain at a GW frequency $\ell F$. The effect of this boost is immediately apparent from Fig.~\ref{fig:hc}, as the difference strain is boosted well-above the sensitivity curve of ET, while the LIGO A+ sensitivity curve is also approached. For even higher eccentricities, this boost is even more pronounced. For $e=0.4$, the difference strain due to tidal dephasing is already above the LIGO sensitivity curve, while for $e=0.6$ and $e=0.8$, the difference strain has significant contributions at GW frequencies even above $f_g$. At these eccentricities, in addition to the boost due to eccentric harmonics we also have a boost due to the increase in $\Delta\Phi^\mathrm{tot}$ from the multiple resonances with different orbital harmonics.

\begin{figure}[!t]
    \centering
    \includegraphics[width=0.48\textwidth]{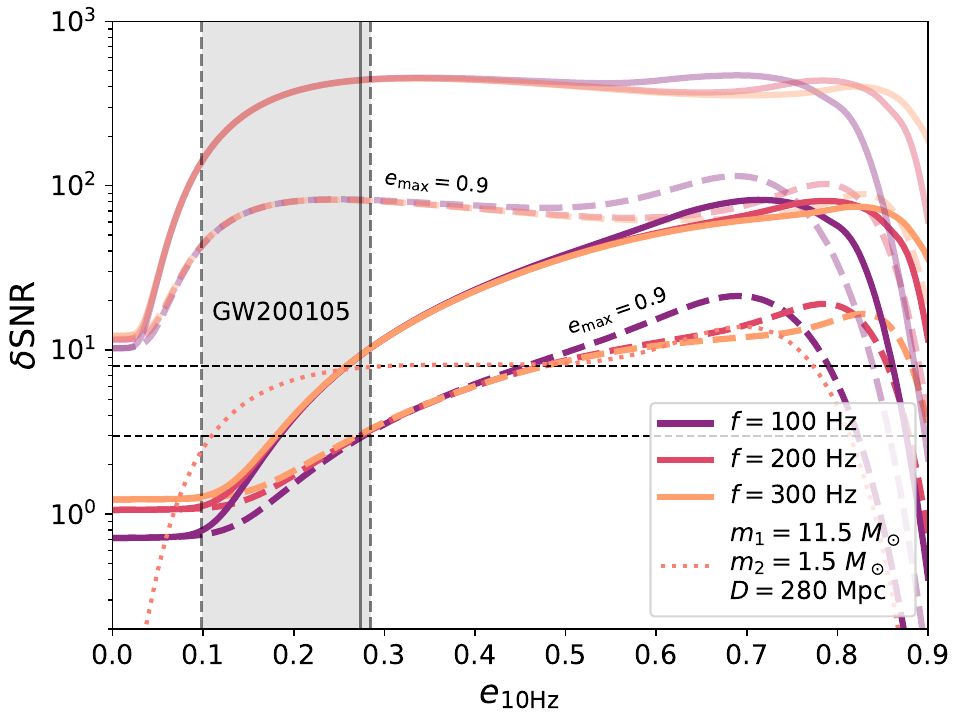}
    \caption{Dependence of the $\delta$SNR of the difference signal on the eccentricity at $10$~Hz for binary NSs with different g-mode frequencies and a tidal coupling $\Lambda_g=0.1$. Dark tones represent results for A+, while semi-transparent lines correspond to ET. Dashed lines were calculated using waveforms truncated at $e_\mathrm{max}=0.9$. The pink dotted line corresponds to observations of NS-BH binaries similar to GW200105 with ET, where the vertical region represents the measured eccentricity, as given by Ref.~\cite{Morras:2025xfu}. Horizontal lines represent $\delta$SNR thresholds of $3$ and $8$. Results for binary NSs assume optimally oriented detectors and binaries at $40$~Mpc distance, where the phase shifts from only one of the NSs were taken into account.}
    \label{fig:dSNR}
\end{figure}

The $\delta$SNR values calculated from these strains are shown in Fig.~\ref{fig:dSNR}, where we used three different values for the g-mode frequency. The solid lines represent results where the full waveform was taken into account, while for the dashed lines the waveforms were truncated below orbital frequencies corresponding to $e(F_\mathrm{min})=0.9$, representing current state-of-the-art eccentric waveform models. Results for both LIGO A+ and ET are shown.
As expected from the strains in Fig.~\ref{fig:hc}, the SNR values start gradually increasing above $e_\mathrm{10Hz}\sim0.1$ for LIGO, while they grow very sharply even below this limit for ET. As mentioned before, this is due to the effect of higher harmonics experiencing a larger dephasing, and has little to do with the multiple epicyclic resonances with the g-mode oscillations that only generate a larger dephasing for $e_\mathrm{10Hz}\gtrsim0.6$ for these g-mode frequencies. Their effect then becomes relevant for larger eccentricities, creating a second peak in the $\delta$SNR curves for ET. Notice also how the SNR values drop for large eccentricities. This is due to several factors. First of all, the total dephasing from g-mode resonances drops for highly-eccentric orbits, as seen in Fig.~\ref{fig:DPhi}, due to the short coherence time of individual resonances. Second, for large eccentricities the SNR from the vacuum-waveforms themselves become small, as can be inferred from the trend in Fig.~\ref{fig:hc}. Finally, for the waveforms truncated above $e\sim0.9$ high eccentricities naturally mean that we are covering an increasingly smaller range in frequency. Another interesting feature of the lines in Fig.~\ref{fig:dSNR} can also be observed. Namely, that lines corresponding to different g-mode frequencies almost completely overlap for eccentricities between $\sim0.2$ and $0.5$. Once again, this can be understood thinking about the reason for the large increase in $\delta$SNR for moderate eccentricities. For such eccentricities the total tidal dephasing is still very similar to the zero-eccentricity case, which is independent of the g-mode frequency in the adopted tidal model. This means that the frequency-domain phase shifts, although generally different at a given orbital frequency, approach the same value at low frequencies. However, since the large increase in the difference strains comes from the phase shifts during the early evolution being transported to higher frequencies by eccentric harmonics, this then makes the $\delta$SNR values approximately equal. Indeed, we see that for low eccentricities, where this eccentric dephasing effect is not relevant, the $\delta$SNR is different for different g-mode frequencies, while the lines also start diverging for large eccentricities, where the total dephasing starts growing in a way that depends on $f_g$.

While the solid lines correspond to the overly-optimistic case of being able to accurately model the waveform up to arbitrary eccentricities, when the more realistic realistic assumption for the waveforms is used, the SNR values are reduced by a factor of $\sim4-5$ for LIGO and a factor of $\sim7-8$ for ET. However, they still represent a significant increase compared to the zero-eccentricity case, reaching the SNR thresholds of $3$ and $8$ at $e_\mathrm{10Hz}\sim0.28$ and $\sim0.46$ for these binary parameters.

Recently, the GW signal from a NS-BH binary inspiral, GW200105, was observed by LVK, which was then later analyzed by multiple groups \cite{Morras:2025xfu,Jan:2025fps,Phukon:2025cky}, who independently found results that support the presence of non-zero orbital eccentricity. Here we also investigated the detectability of g-mode resonance induced in the NS component of binaries with similar parameters. We adopted the results of Ref.~\cite{Morras:2025xfu}, who inferred the binary parameters to be $m_1\approx11.5~M_\odot$, $m_2\approx1.5~M_\odot$, while the eccentricity of the binary was inferred to be $e_\mathrm{20Hz}=0.145^{+0.007}_{-0.097}$ with $90\%$ confidence, which corresponds to $e_\mathrm{10Hz}=0.274^{+0.011}_{-0.176}$. The results shown on Fig.~\ref{fig:dSNR} are for the ideal case of accurate modeling of the entire waveform and observation with ET. We see that the SNR values are much below for this system than for binary NSs at $40$~Mpc, although for the inferred eccentricity they are comparable to the $\delta$SNR from a circular binary NS. This decrease is due to two main factors. One of them is the binary being much further away, while the other is the large primary mass, which makes the tidal forces much weaker at a given orbital frequency. However, even though eccentric binary NSs are yet to be observed, the existence of eccentric BH-NS inspirals might provide an alternative way of testing the properties of g-modes in the future.

\begin{figure*}[!t]
    \centering
    \includegraphics[width=0.497\textwidth]{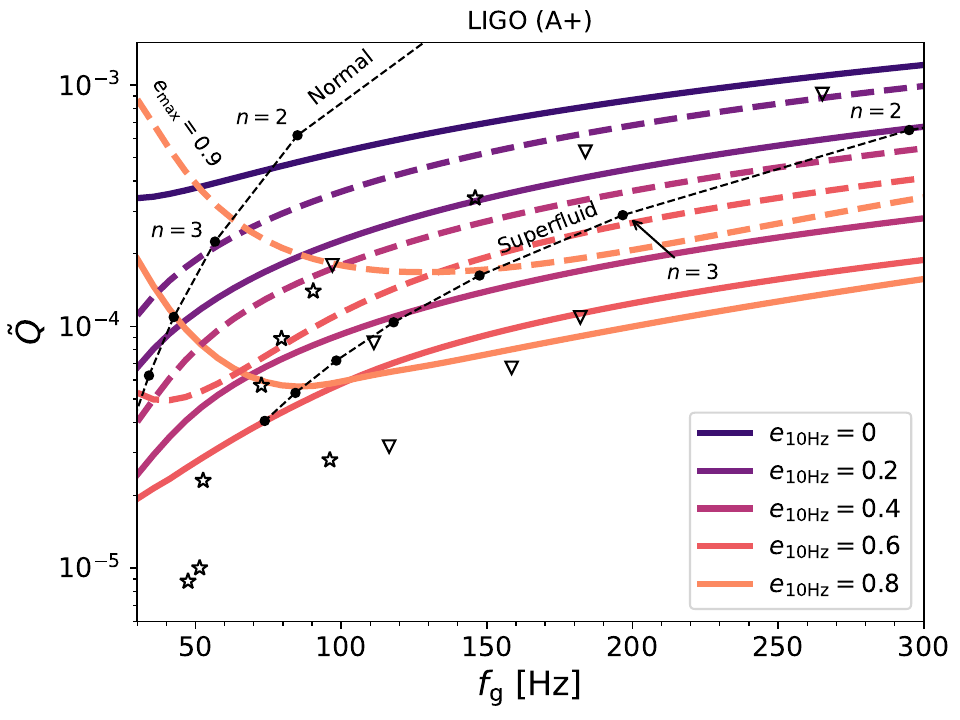}
    \includegraphics[width=0.497\textwidth]{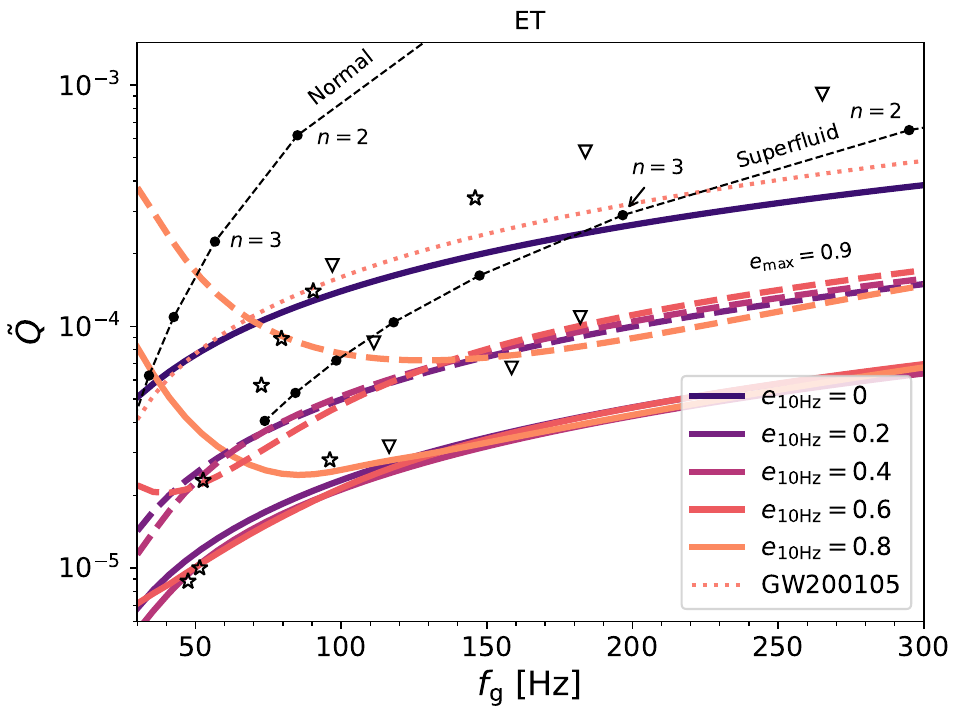}
    \caption{Upper bounds on the dimensionless overlap-integral of g-mode tides, using $\delta\mathrm{SNR} \geq 3$, as a function of the g-mode frequency for different eccentricities, for the A+ detector (left panel) and for ET (right panel). Dashed lines were calculated using waveforms truncated at $e_\mathrm{max}=0.9$. The pink dotted line in the right panel corresponds to the observation of a NS-BH binary similar to GW200105. Black data points represent predictions of microphysical calculations using different equations of state. Red dots connected by dashed lines correspond to results on normal and superfluid nuclear matter from Refs.~\cite{Yu:2016ltf,Ho:2023shr}, while red stars and triangles represent results from Refs.~\cite{Counsell:2024pua} and \cite{Lai:1993di}, respectively. Results for binary NSs assume optimally oriented detectors and binaries at $40$~Mpc distance, where the phase shifts from only one of the NSs were taken into account.}
    \label{fig:Qf}
\end{figure*}

Finally we study how these $\delta$SNR values translate to constraints on the tidal parameters, namely the g-mode frequency and the overlap integral $\tilde{Q}$. For this we use the condition that tidal parameters with $\delta\mathrm{SNR}>3$ would be detectable, and hence parameters producing larger SNRs could be constrained by observations. The constraints from binary NSs with different eccentricities are shown in Fig.~\ref{fig:Qf}, for LIGO A+ and ET.
Results for g-mode properties from various equations of state are also shown for normal and superfluid nuclear matter from Refs.~\cite{Yu:2016ltf,Ho:2023shr} with circles connected by dashed lines, with triangles from Ref.~\cite{Counsell:2024pua}, and with stars from Ref.~\cite{Lai:1993di}. These show results not only for first g-modes, but for higher modes as well ($n>1$). For idealistic waveform models the constraints could be substantially improved even for binary NSs with moderate eccentricities of $e_\mathrm{10Hz}\sim0.2$ observed by LIGO, while with $e_\mathrm{10Hz}\sim0.4$ current state-of-the-art eccentric waveforms could also reach this level. With ET the situation is naturally even better, with the possibility of constraining the strongest g-modes of essentially all relevant models with a complete eccentric model and binary NSs with $e_\mathrm{10Hz}\gtrsim0.2$. Robust constraints could also be put using only current waveform models. Notice how due to the flat $\delta$SNR curves in Fig.~\ref{fig:dSNR} for ET, there is no additional gain in terms of these constraints by increasing the eccentricity of the binaries beyond $0.2$. Moreover, for very eccentric binaries we even loose some of our sensitivity, especially at low frequencies, where binaries with $e\sim0.8$ perform worse than circular binaries. Finally, note that due to the fact that $\Delta\Phi\propto\Lambda_\alpha\propto\tilde{Q}_\alpha^2$, by increasing the $\delta$SNR of a GW signal by two orders of magnitude, we only obtain constraints on $\tilde{Q}$ that are improved by a single order of magnitude.

\section{Discussion}
\label{sec:conclusion}

In this paper we demonstrated that moderate non-vanishing eccentricities can significantly increase the detectability of low-frequency NS tides. We provided analytic formulas for calculating the orbital phase shift for arbitrary eccentricities and g-mode frequencies and validated them through comparison with results from direct numerical integration. We have found that the amplification of the phase shift comes from two main sources. (i) For finite eccentricities, quadrupole g-mode tides will resonate with not just the $\ell=2$ harmonic of the orbit, but with multiple higher harmonics with $\ell>2$, and thus all of these resonance add up to create a larger phase shift. (ii) Due to the eccentric orbit, phase shifts corresponding to the early phase of the evolution are being transported up to frequencies within the sensitive band of GW detectors by all the higher harmonics present in the waveform. While the first effect only has a significant contribution at high eccentricities of $e_\mathrm{10Hz}\gtrsim0.5$, the second effect contributes to the amplification already at $e_\mathrm{10Hz}\gtrsim0.1$. We have shown that the increase in detectability of g-mode resonances due to these two effects could enable putting constraints on tidal parameters that are an order of magnitude better than what we would get from circular binaries, through observations of eccentric binary NSs with moderate eccentricities, $e_\mathrm{10Hz}\sim0.2-0.4$. This potentially enables putting robust constraints on g-mode parameters, and consequently on properties of NS matter a high densities, already with current GW detectors. Note that while the results in this paper were derived for resonances with the first g-modes in NSs, the same framework could be applied to other low-frequency modes as well with a frequency $\mathcal{O}(100)$~Hz, such as interface modes. Also note that we only accounted for g-mode resonances in one of the binary components, and used the expectation value of the total dephasing to calculate our results. For binary NSs both compact objects will contribute to the total phase shift through g-modes with potentially different properties. These contributions will then simply add up, creating a larger overall effect. Moreover, the total phase shift can be smaller but also significantly larger as well than the expectation values derived here.

Our results are based on the $\delta$SNR criterion \cite{Kocsis:2011dr}. While this provides a good idea about whether there is a possibility for an effect to be observed, it does not take into account degeneracies with other orbital parameters. Even weak degeneracies in general make it more difficult to measure small additional effects, such as g-mode resonances. We expect, however, that due to the distinct frequency dependence of the phase shifts produced by g-mode resonances, corresponding to almost instantaneous energy extractions from the orbit, degeneracies with variations in the orbital parameters, which induce smooth changes in the GW phase, will be minimal. Nonetheless, in order to make a firm statement about the increase in detectability of g-mode tides, one needs to carry out a full parameter inference, including the effect of g-mode resonances. This problem is reserved for a follow-up study.

Throughout the paper we made a number of assumptions. First of all, we carried out our calculations for g-mode resonances in a Newtonian manner, while also applying a Newtonian waveform model. We expect that the former, with the proper general relativistic tidal parameters, is a good approximation, since these resonances occur at binary separations where post-Newtonian effects are relatively weak. However, the inclusion of a higher-order post-Newtonian waveform would certainly affect our results, most importantly, due to the precession of the orbit, which we completely neglected in this study. However, no significant changes are expected in our conclusions once these post-Newtonian effects are included in the waveform model.

We also assumed that only a single low-frequency mode contributes to the phase shift of the orbit. This is true for simple equations of state without strong phase transitions, and non-spinning NSs, where $n>1$ g-modes are suppressed compared to the first g-mode. However, phase transitions and spins can give rise to additional tidal modes that can potentially have comparable strengths to these g-modes \cite[see e.g.][]{Counsell:2025hcv,Zhu:2022pja,Lau:2020bfq,Rodriguez:2025oes}. Additionally, environmental effects, such as gaseous drag or line-of-sight acceleration due to a tertiary companion could also appear in the low-frequency part of the GW signal \cite[e.g.][]{Samsing:2024syt,Hendriks:2024gpp,Zwick:2025wkt,Zwick:2025qzv,Saini:2025ncj}. These additional effects would make parameter inference more complicated and potentially bias the inferred properties of g-mode tides.

Additionally, we neglected the spins of NSs during our derivations. For f-modes, with a frequency of $\mathcal{O}(10^3)$~Hz, NS spins of a maximum of a few hundred Hz only introduce a small second-order effect, in addition to shifting the resonance frequency of the modes. However, rotation could potentially have a significant effect on g-modes, the frequencies of which is comparable to the rotation frequency of highly spinning NSs. First of all, rotation will split resonances with different $m$, and amplify or deamplify the phase shift from resonances with prograde or retrograde modes. Moreover, it will also give rise to additional inertial modes, that could complicate the tidal spectrum even further. For higher rotation frequencies some modes can even become unstable and eventually change the spin of the NS through non-linear coupling with the bulk matter. For such unstable modes, bulk and shear viscosity also plays an important role, which could effectively suppress these instabilities. Hence, resonant excitations of g-modes in moderately-rotating NSs becomes a fairly complex problem, which we did not attempt to pursue in this paper. However, we expect the effect of finite eccentricities to be the same, only coupled with a much more complicated tidal spectrum. The impact of rotation on NS tidal modes had been studied in great detail and we refer the reader to e.g. Refs.~\cite{Andersson:1997xt,Ho:1998hq,Andersson:1998ze,Xu:2017hqo,Poisson:2020eki,Pnigouras:2022zpx}, for further reading.

Finally, our study assumes that a reasonably large subpopulation of eccentric binary NSs or NS-BH binaries exist, which enable the detection of at least a single significant event with LVK or ET. Due to mass segregation of heavier BHs towards the core of globular clusters, NSs encounter less dynamical interactions. Therefore the rates of BNSs in globular clusters is expected to be relatively low~\cite{Bae:2013fna}. The observed Galactic BNSs are likely to have eccentricity $\mathcal{O}(10^{-5})$ at $10$ Hz GW frequency. Nevertheless, various formation mechanism can gives rise to eccentric binaries in the LVK frequency band. Kozai-Lidov mechanisms in hierarchical triple systems can lead to the larger residual eccentricity of BNS systems~\cite{PhysRevLett.111.061106}. Binary neutron star mergers in young star clusters~\cite {Ziosi:2014sra}, dynamical interactions of BNSs in globular clusters~\cite{2013MNRAS.428.3618C}, BNS and NS-BH mergers in galactic nuclei~\cite{Petrovich:2017otm} are the potential formation mechanisms that can ehnance the merger rates of eccentric NS binaries. Third-generation detectors will detect $\mathcal{O}(10^5)$ BNS and NS-BH mergers every year~\cite{Gupta:2023lga}. Even though a comprehensive population synthesis study for these channels would be needed to infer the detection rate of NS binaries with measurable eccentricities, the observation of GW200105, corresponding to an NS-BH merger with non-zero eccentricity, alone provides grounds for expectations for several eccentric binary NSs or NS-BH binaries to be observed with improved current or next-generation GW detectors.

\begin{acknowledgments}

The authors would like to thank Bence Kocsis for helpful discussions and useful comments on the manuscript.
J.T. acknowledges support from the Horizon Europe research and innovation programs under the Marie Sk\l{}odowska-Curie grant agreement no. 101203883, and support from the Alexander von Humboldt Foundation under the project no. 1240213 - HFST-P. L.Z. is supported by the European Union’s Horizon 2024 research and innovation program under the Marie Sklodowska-Curie grant agreement No. 101208914. This work was supported by the ERC Starting Grant No. 121817–BlackHoleMergs led by Johan Samsing, and by the Villum Fonden grant No. 29466. The Center of Gravity is a Center of Excellence funded by the Danish National Research Foundation under grant No. 184.

\end{acknowledgments}


\bibliography{Lowfreq}{}


\end{document}